\documentclass[journal]{IEEEtran}
\newcommand{\userdefinelength}{0.75\linewidth}

\usepackage{kotex}
\newcommand*\pFq[2]{{}_{#1}F_{#2}}

\ifCLASSINFOpdf
\else
\fi
\usepackage[labelformat=simple]{subcaption}

\usepackage{graphics,amssymb,graphicx, theorem,amsmath,amsfonts,epsfig,float,times,multirow, setspace, latexsym,psfrag,cite,ulem, algpseudocode, algorithm, amsmath}

\usepackage{bbm}

\usepackage{xcolor}
\usepackage{soul}

\usepackage{graphicx}
\usepackage{lipsum}
\usepackage{siunitx}
\usepackage{placeins}

\definecolor{LightBlue}{rgb}{0.75,0.936,1.00}
\sethlcolor{LightBlue}
\definecolor{LightCyan}{rgb}{0.88,1,1}

\allowdisplaybreaks

\newcommand{\be}{\begin{equation}}
\newcommand{\ee}{\end{equation}}
\newcommand{\ben}{\begin{eqnarray}}
\newcommand{\een}{\end{eqnarray}}

\begin{document}

\title{Phase-Shift Design and Channel Modeling for Focused Beams in IRS-Assisted FSO Systems}

\author{Jung-Hoon Noh~\IEEEmembership{Member,~IEEE} and
Byungju Lee~\IEEEmembership{Member,~IEEE}\vspace{-1.5\baselineskip}}

\vspace{-100pt}
\date{\today}
\maketitle 

\begin{abstract}
Interest in free-space optics (FSO) is rapidly growing as a potential solution for the backhaul of next-generation mobile or low-orbit satellite communications. 
Various techniques have been suggested for employing an intelligent reflecting surface (IRS) in FSO systems, such as anomalous reflection, power amplification, and beam splitting. 
It is possible to deliver more power to the receiver (Rx) by collimating or focusing the reflected beam at the Rx lens. 
In this study, we propose a phase-shift design of an IRS for beam focusing. 
In addition, we propose a new pointing error model and an outage performance analysis applicable when the beam width is comparable to or less than the aperture size of the Rx. 
The analytical results are validated by Monte Carlo simulations. This study provides essential preliminary results for future researches that assume a focused beam in FSO systems. 
\end{abstract}


\vspace{-5pt}

\section{Introduction}
\label{sec:intro}

Optical wireless systems, such as free-space optical (FSO) systems, are a potential technology for high-data-rate transmission with license-free operation over a wide range of bandwidth in the optical spectrum \cite{Khalighi:2014}. 
In contrast to radio frequency (RF), FSO systems are immune to electromagnetic interference and have been considered a cost-effective solution for terrestrial backhaul/fronthaul wireless applications for 5G and beyond 5G networks. Recently, the use of an intelligent reflecting surface (IRS) in FSO systems has attracted significant research interest  [2-12]. 
An IRS is an electromagnetic device with electronically controllable characteristics. Various techniques can be used to manipulate the amplitude, phase, and polarization of the incoming signal using IRS. For example, a power-amplifying IRS module was proposed in \cite{Ndjiongue:2021-1}.
The authors in \cite{Ajam:2022} and \cite{Wang:2020} introduced power splitting to multiple receivers (Rx) using IRS.

Anomalous reflection is a fundamental technique which freely adjusts the direction of the reflected beam by manipulating the phase of the incoming beam over the surface of the IRS.
Through anomalous reflection using IRS, we can alleviate the line-of-sight (LOS) requirements, which are essential for establishing FSO links. 
For the anomalous reflection of an IRS-assisted FSO system,  
\cite{Najafi:2021} proposed a phase-shift design and pointing error investigation. 
Moreover, analytical expressions for the end-to-end channel model \cite{Ajam:2021} and various performance metrics \cite{Bosu:2019,Chapala:2021,Yang:2020,Ndjiongue:2021-2,Jia:2020} were introduced in which both atmospheric turbulence and pointing errors were considered.


Furthermore, there is increasing interest in techniques that not only control the direction of the beam, but also its size  \cite{Najafi:2021,Jamali:2021}. 
Using the IRS, we can adjust the convergence of the reflected beam at the Rx. However, research using IRS to create a focused beam is still in its infancy; not just the IRS phase-shift designs, but research into the pointing error for the focused beam remains an open problem \cite{Jamali:2021}. 
While the focused beam using the IRS system may deliver greater power to the Rx than a general reflected Gaussian beam under perfect beam tracking, its performance can be more sensitive to pointing errors. 

In this study, we first propose a phase-shift design of the IRS that can freely adjust the beam width as well as the direction of the beam. Then, we present a pointing error model for the case where the beam width is similar to, or smaller than, the diameter of the Rx lens. It should be noted that the conventional Gaussian form pointing error model can be applied only if the beam width is more than twice the diameter of the Rx aperture \cite{Farid:2007}. 
Furthermore, using the newly proposed pointing error model, the outage probability of the system was investigated when the effects of the pointing error and atmospheric loss were simultaneously considered. The proposed phase-shift design and pointing error model were verified using simulation results.

Finally, we discuss how the outage performance is affected by the beam width, atmospheric loss, transmitted power, and pointing error. These factors are intricately intertwined, making it difficult for system designers to determine the system parameters. However, the presented mathematical modeling can facilitate performance analysis in terms of these factors for the focused beam. In addition, it offers insight into system design. The size of the beam can be determined to achieve the best performance in a combination of several channel influences.
\vspace{-5pt}

\section{System Model}
\label{sec:sys_model}

\subsection{Power Density of Gaussian Beam}

\begin{figure}
\centering
  \includegraphics[width=0.8\linewidth]{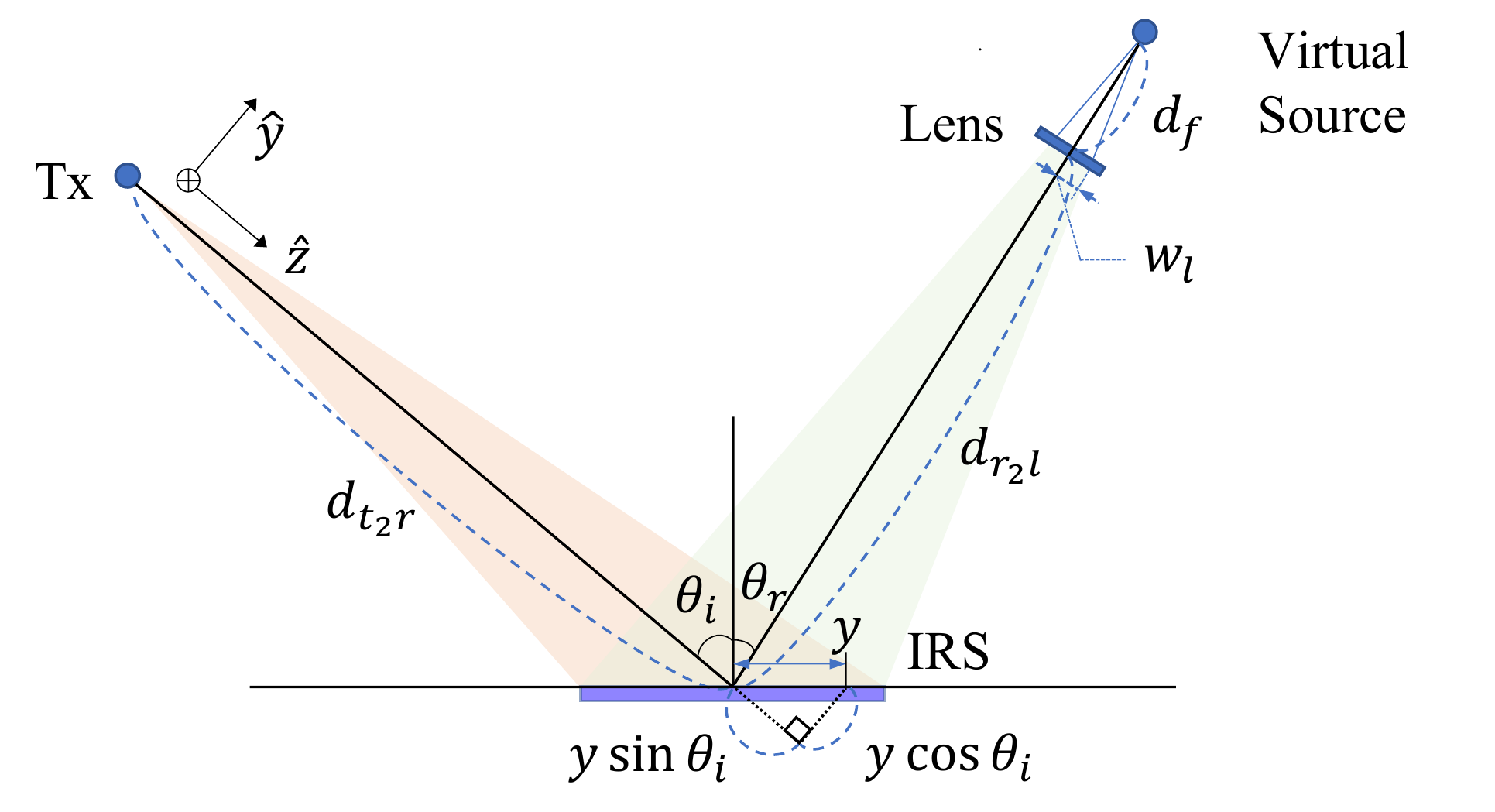}
  \captionof{figure}{Illustration of focused beam shaping using IRS}
  \label{fig:IRS_control_design}
  \vspace{-15pt}
\end{figure}

First, we present the electric field and power density distributions of the transmitted Gaussian beam on the IRS. 
We consider the two-dimensional (2D) coordinate system of the $yz$-plane, where the IRS exists on the $y$-axis.
Without loss of generality, it is assumed that both the centers of the IRS and transmitted beam are at the origin, as illustrated in Fig. \ref{fig:IRS_control_design}.
The lengths of the IRS  line and  the Rx lens lines are denoted as $2a_r$ and $2 a_l$, respectively. $\theta_i$ and $\theta_r$ are the incident and reflection angles, respectively, at the IRS.

Next, we define the electric field and power density of the Gaussian beam in the positive $\hat{z}$ direction from the transmitter using $\hat{y}\hat{z}$-coordinates. 
The electric field is expressed as follows: 
\begin{align}
&\hspace {-.5pc} E(\hat{y},\hat {z} ; w_0)=E_{0}
\sqrt{\frac {w_{0}}{w(\hat{z},w_0)}}
\nonumber \\&\qquad \times 
\exp \left(-\left({\frac {\hat{y}^{2}}{w^{2}(\hat {z},w_0)}} + j\psi(\hat{y},\hat{z},w_{0})\right)\right)
\end{align}
where phase
\begin{equation} \psi _{\mathrm {G}}(\hat{y},\hat {z},w_{0})=-k\hat {z}-k\frac {\hat{y}^{2}}{2R(\hat {z},w_0)}+\zeta (\hat {z}),
\end{equation}
$\hat{y}$ is the axis perpendicular to $\hat{z}$, 
$E_0$ denotes the electric field at the origin, 
$w_0$ denotes the beam waist radius, $k=2 \pi \lambda$ is the wave number,  $\lambda$ is the optical wavelength, $R(\hat{z},w_0)=\hat{z} (1+(z_0/\hat{z})^2)$ is the curvature radius of the beam’s wavefront at $\hat{z}$, 
$z_0 = \pi w_0^2/\lambda$ is the Rayleigh range, $\zeta(\hat{z}) = \tan^{-1}\left( \hat{z}/z_0 \right)$ is the Gouy phase, and $w(\hat{z},w_0) = w_0 \sqrt{1+\left(\hat{z}/z_0 \right)^2}$ is the beam width,  where the intensity values fall to $1/e^2$ of their axial values on the $\hat{z}$ plane.  
In addition, the power density distribution of the Gaussian beam 
$I(\cdot) = | E(\cdot) |^2$ is given by
\be
I(\hat{y},\hat{z} ; w_0) = \frac{2}{\sqrt{\pi} w(\hat{z},w_0)} \exp \left( -\frac{2\hat{y}^2}{w^2(\hat{z},w_0)} \right).   
\ee

Then, we express the electric field across the IRS as
$E(y \cos \theta_i, d_{t2r} + y \sin \theta_i ; w_0)$, and approximate the power density distribution across the IRS as follows: 
\begin{multline}
I_{irs}(y , d_{t_2r} ; \theta_i, w_0 ) \\
\approx \frac{\sqrt{2} \cos \theta_i}{\sqrt{\pi} w(d_{t_2r},w_0) } 
\exp \left( - \frac{2 cos^2 \theta_i y^2}{ w( d_{t_2r}, w_0) } \right),
\label{eq:power_dist_IRS} 
\end{multline}
where the approximation arises when $d_{t2r} \gg y \sin \theta_i$.

\vspace{-2pt}
\subsection{Channel Model}

Next, we define the FSO communication link. This study assumes that 
the intensity modulation and direct detection are employed where the on-off keying~(OOK) is used as the modulation scheme. Laser beams propagate through a turbulence channel with additive white Gaussian noise (AWGN) in the presence of pointing errors. The received optical signal is then converted into an electrical signal at the photodetector. 
The received signal, $y$, can be expressed as $y = \eta h x + n$, where $x$ is the transmitted signal, $\eta$ is the effective photo-electric conversion ratio, $h$ is the channel gain, and $n$ is the signal-independent AWGN with a variance of $N_0$. 

The channel gain, $h$, can be modeled as $h = h_l h_a h_p$,  where $h_l$ represents the attenuation caused by the scattering and absorption of particles in the air, $h_a$ is the atmospheric fading loss factor, and $h_p$ is the pointing error loss factor. Note that $h_a$ and $h_p$ are both independent random variables, whereas $h_l$ is a constant given by $h_l = \exp\left(-\sigma z\right)$ where $\sigma$ is attenuation coefficient and $z$ is link distance. 

Becuase the coherence time of FSO channels 
is considerably longer than the typical bit interval, 
owing to the high data rates, the FSO channel is commonly modeled as a slow-fading channel, where the outage probability becomes a relevant metric in performance evaluation. 
Given the rate requirement $R_0$, the outage probability is expressed as $P_{out} = \mbox{Prob}( h < h_0)$  
where $h_0 = \sqrt{ \left(2^{R_0}-1 \right) \sigma_n^2/(\sqrt{2}P_t R_0)^2 }$,
and $P_t$ is the average transmitted optical power. 

\section{Phase-Shift Design of IRS}
\label{sec:irs_ctrl}

This section presents the IRS phase-shift design for creating a reflected beam focused on the Rx. 
By manipulating the phase of the reflected beam over the IRS surface, we can control the convergence of the beam at the Rx. 
First, we consider the case when the reflected beam is completely focused at the center of the Rx. 
A focused beam can be created by allowing the beam reflected from the IRS to have a phase profile of the virtual beam propagating from the center of the Rx with the opposite direction. 

Assuming a mirror-assisted system, we consider the virtual beam that comes 
from the Rx lens, whose location is specified by $\theta_r$ and $d_{r_2l}$. 
It is assumed that the virtual beam has the same power density distribution as the original beam from the transmitter~(Tx) on the IRS, as follows:
\be
I_{irs}(y, d_{t_2r}  ; \theta_i, w_0 ) 
= I_{irs}(y , d_{r_2l} ; \theta_r, \tilde{w}_0 ),
\ee
where $\tilde{w}_0$ is the beam waist of the virtual beam. As the solution of the equation
$w( d_{r_2l}, \tilde{w}_0 ) = w_{eq,\theta_r}$, we obtain 
\be
\tilde{w}_0 = \sqrt{ \frac{w_{eq,r}^2}{2} - \frac{1}{2\pi} 
\sqrt{ \pi^2 w_{eq,r}^4 - 4 d_{r_2l}^2 \lambda^2 } }, 
\label{eq:w_0_tilde}
\ee
where $w_{eq,\theta_r} =  w_{irs} \cos \theta_r$ and
$w_{irs} =  w( d_{t_2r}, w_0)/\cos \theta_i$ is the  width of the transmitted beam on the IRS.  

Next, we construct the phase-shift profile $\Delta \psi( y )$ which is 
added to the phase of reflected beam across the IRS.  
It is built by adding the phase of the virtual beam with reversed direction, that is 
$\psi\left( y \cos \theta_r, -(d_{r_2l} +  y \sin \theta_r) , \tilde{w}_0 \right)$, and
subtracting the phase of the incoming beam, 
that is $\psi( y \cos \theta_i, d_{t_2r} +  y \sin \theta_i , {w}_0 )$.
The resulting phase-shift profile across the IRS becomes
\begin{multline}
\Delta \psi( y | \theta_i , \theta_r, w_0 ,\tilde{w}_0 ) = \pi 
- \psi\big( y \cos \theta_i, d_{t_2r} +  y \sin \theta_i , {w}_0 \big) \\
+ \psi\big( y \cos \theta_r, -(d_{r_2l} +  y \sin \theta_r) , \tilde{w}_0 \big).
\label{eq:phase_shift_prof}
\end{multline}
Eventually, the reflected beam is focused at a distance $d_{r_2l}$ with the beam width of $\tilde{w}_0$.

Furthermore, we consider the case 
when the beam width at the Rx is larger than $\tilde{w}_0$ and smaller than $w_{irs}$. 
In this case, we assume that reflected beam is focused at a greater distance by $d_f$ than the Rx, where the virtual beam source is located. 
Then, we must find the virtual beam whose width is $w_{l}$ at the RX lens of distance $d_f$, and $w_{irs}$ at the IRS of distance $d_{r_2l} + d_f$. 
To obtain the exact $\tilde{w}_0$ of such a beam, the following equations must be solved: 
\be
\frac{w( d_{r_2l} + d_f , \tilde{w}_0 )}{cos \theta_r} = \frac{w( d_{t_2r}, w_0)}{cos \theta_i} \, \mbox{ and } \, w(d_f,\tilde{w}_0) = w_l.
\label{eq:w_set_eq}
\ee
However, the analytical solution of \eqref{eq:w_set_eq} is very complicated, because it requires the roots of a general cubic equation to obtain $\tilde{w}_0$. 
Instead of an exact analytical solution, we present an approximated solution. 
Assuming the Rx is at a great distance from the IRS, i.e., 
$d_{r2l} \gg z_R$,
$\tilde{w}_0 $ can be approximated as follows \cite{Goodman:2005}:
\be
\tilde{w}_0 \approx \frac{ \lambda d_{r2l}}{\pi ( w_{eq,\theta_r} - w_l)} 
\label{eq:w_hat_for_wset}
\ee
Note that \eqref{eq:w_hat_for_wset} can be verified by \eqref{eq:w_set_eq}.

Further, we compare the power density distributions of reflected beam at the Rx, which are obtained by two different method: the geometric optics and the Huygen-Fresnel principle. 
The power density of the geometric optics was obtained by the beam propagated from virtual beam source to the Rx, at a distance $d_f$ from the source. 
The Huygens-Fresnel principle treats every point on the IRS as a secondary point source emitting a spherical wave \cite{Goodman:2005}. Therefore, the electric field on the Rx is the superposition of the waves originating from all secondary sources. This is expressed as follows:
\begin{align*}
&\hspace {-.5pc} E_{\mathrm{ Rx}} (\tilde{y}) = \frac {\varsigma }{j \sqrt{\lambda}} \int_{-a_r}^{a_r} E_{\mathrm{ irs}} (y) \frac {\exp (jk|\mathbf {r}_{0}-\mathbf {r}_{1}|)}{\sqrt{|\mathbf {r}_{0}-\mathbf {r}_{1}|}} \\&\qquad \qquad \qquad \qquad \qquad \qquad \times \exp (j\Delta \psi (y)) dy, \tag{9}
\end{align*}
where $\varsigma = \sqrt{\cos \theta_i/\cos \theta_r}$, 
$\mathbf{r}_0 =
\big[\begin{smallmatrix}
  \cos \theta_r & \sin \theta_r \\
  -\sin \theta_r & \cos \theta_r 
\end{smallmatrix}\big]
\big[\begin{smallmatrix}
  \tilde{y} \\
  d_{r_2l} 
\end{smallmatrix}\big]$, $\mathbf{r}_1 = 
\big[\begin{smallmatrix}
  y \\
  0 
\end{smallmatrix}\big]$, and 
$\Delta \psi (y)$ is the phase-shift profile across the IRS, given by \eqref{eq:phase_shift_prof}.
\begin{figure}
\centering
  \includegraphics[width=\userdefinelength]{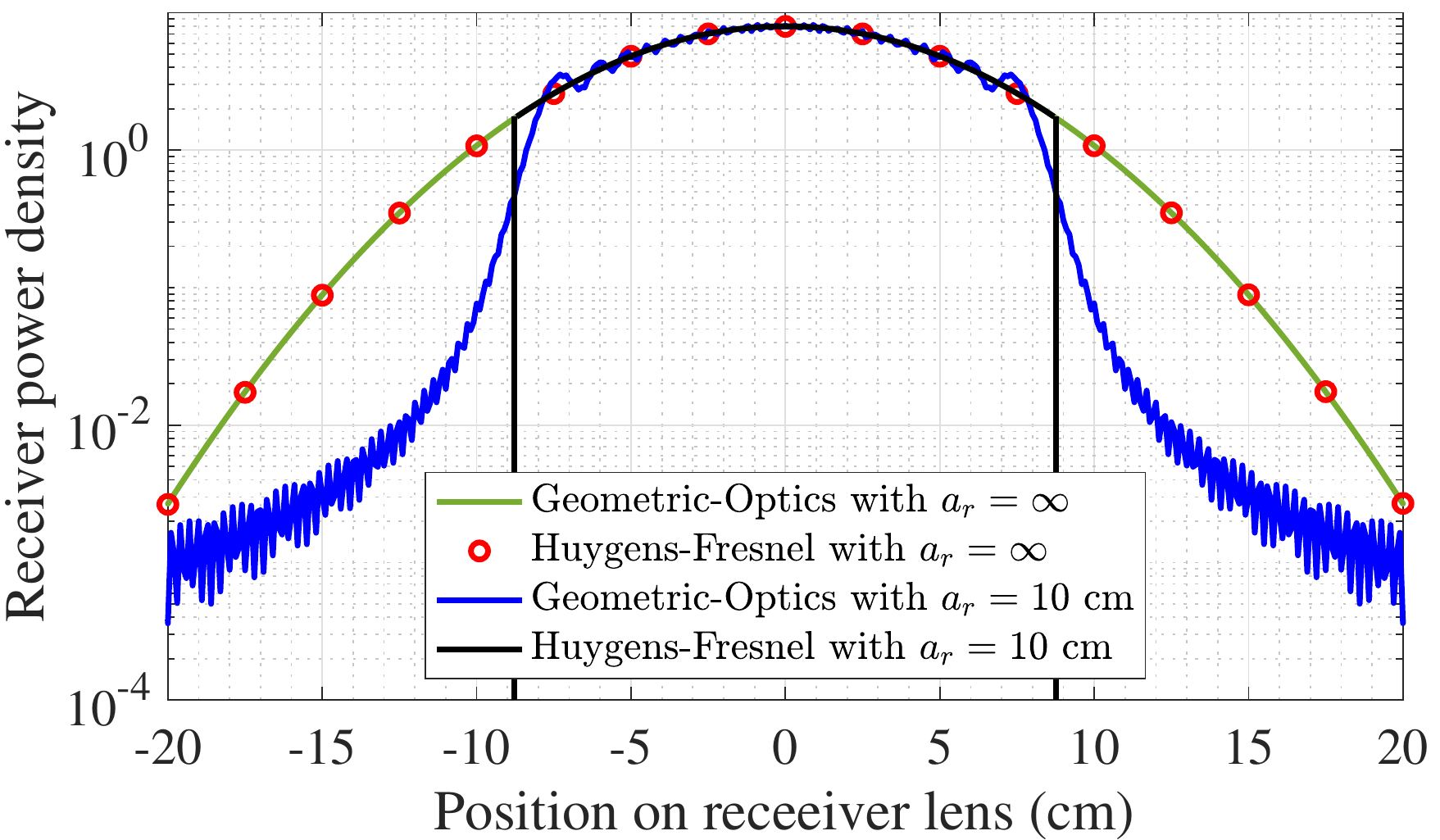}
  \captionof{figure}{Power density distribution on Rx lens with $\theta_i = \pi/3$, 
  $\theta_r =\pi/6$, $w_0= 1$ mm, $a_l=10$ cm, and $a_r \in \{10, \infty\}$.}
  \label{fig:power_density_dist_at_RX}
  \vspace{-20pt}
\end{figure}

The results of this comparison are presented in Fig. \ref{fig:power_density_dist_at_RX}. 
The received power density computed from the Huygens-Fresnel integration matches well with the geometric optics computation. 
Thus, the proposed phase-shift design for focusing beams is verified. 
Expanding this result to three dimensions can be achieved using a rotated astigmatic Gaussian beam. 
The results of 2D case in the $yz$ plane can be repeated for the $xz$ plane, and eventually extended to the rotated astigmatic Gaussian beam for the three-dimensional~(3D) case. 
Further details of extension to 3D system are demonstrated in \cite{Najafi:2021}, and are not covered in this paper owing to the limitation of the length.

So far, we have introduced a method to focus the beam at the Rx using the IRS. With the proposed phase-shift design, we can concentrate the transmitted power on the Rx using the IRS. However, the focused beam is more sensitive to pointing errors. The following section examines the effect of the pointing error when the beam is focused, that is, when the beam width is similar to or smaller than the Rx size.

\section{Pointing Error Modeling}

This section examines the effect of pointing errors on the quality of the FSO channels when the beam is focused on a size similar to or smaller than the Rx aperture. 
Note that the laser beam reflected from the IRS becomes an astigmatic Gaussian beam at the Rx lens in the 3D space. 
However, we considered only a general Gaussian beam. This is because the main purpose of the pointing error modeling is to analyze the influence of pointing errors according to the size of the beam rather than its shape. Based on the results of the general Gaussian beam, we can expand it to an astigmatic Gaussian beam in future work. 

\vspace{-10pt}

\subsection{Channel Coefficient Modeling} 

First, we introduce the pointing error model and then derive the probability density function~(PDF) for two cases according to the beam width.
The first case is when the beam width is similar to the radius of Rx lens. This can be approximated using the Gaussian error function, i.e., $\mbox{erf}(\cdot)$ as follows:
\begin{align}
\hspace {-.5pc} 
h_p (u ; w_l, a_l)
 &= \int_{-a}^{a} \int_{-\sqrt{a_l^2-x^2}}^{\sqrt{a_l^2-x^2}}
\frac{2}{\pi w_l^2} \nonumber \\
&\qquad  \qquad \times \exp{\left( -2\frac{(x-u)^2+y^2}{w_l^2} \right)} dy dx  \label{eq:exact_h_p} \\
\,&\stackrel{(a)}{\approx} \int_{-\infty}^{\frac{\sqrt{\pi}a_l}{2}} \frac{\sqrt{2} E}{\sqrt{\pi w_l^2}} 
\exp{\left(\frac{-2(x-u)^2}{w_l^2}\right)} dx \nonumber \\
\,&= \frac{E}{2} \left( \mbox{erf} \left( \frac{\sqrt{2}}{w_l}\left(\frac{\sqrt{\pi}}{2}a_l-|u|\right)  \right) +1 \right), 
\label{eq:erf_appro_h_p}
\end{align}
where the beam propagates in the positive $z$ direction, the Rx lens is located on the $x\-y$-plane, $E$ is derived from the error function where $E = {\rm{erf}} \left( \sqrt{\pi}a_l /(\sqrt{2}w_l) \right)$, $w_l$ is the beam width at the Rx lens, and $u$ is the displacement of beam from the center of the Rx.  
In \eqref{eq:erf_appro_h_p}, (a) follows from 
that the power of the beam outside the Rx lens is negligibly small when the beam width is similar to or smaller than the radius of Rx aperture.

Assuming a Gaussian fluctuation of the center of beam on both the horizontal and vertical axes of the Rx, $u$ follows a Rayleigh distribution. 
The PDF of the pointing error loss, $h_p$, is  then given by
\begin{equation}
f_{h_p} (h_p) = \sqrt{\frac{\pi}{2}} w_l \lambda(h_p) 
\exp{\left( v(h_p)^2-\frac{\lambda(h_p)^2}{2 \sigma^2}  \right)}
\label{eq:hp_pdf_erf}
\end{equation}
where $\lambda(h_p) = \frac{\pi}{2}a_l-\frac{w_l}{\sqrt{2}} v(h_p)$, $v(h_p) = \mbox{erf}^{-1}\left( \frac{2}{E} h_p - 1 \right)$,
$\mbox{erf}^{-1}(\cdot)$ is the inverse of the $\mbox{erf}(\cdot)$, $0 \leq h_p \leq A_0$, and $A_0~=\frac{E(E+1)}{2}$ is the fraction of the collected power at $r=0$.

Meanwhile, integrating \eqref{eq:hp_pdf_erf} to obtain a performance metric such as a outage probability may be limited in cases where $w_l \ll a_l$.
Note that for $w_l \ll a_l$, $E \approx 1$ and $A_0 \approx 1$. 
Then, $f_{h_p}(A_0)$ becomes infinite, and  \eqref{eq:hp_pdf_erf} cannot be integrated from $0$ to $A_0$. 
Thus, we introduce an another pointing error model applicable for the case of $w_l \ll a_l$ as follows:
\be
h_p(u;a_l) = \mathbbm{1}_{A} (u)
\label{eq:Indic_hp}
\ee
where $A=[-a_l, a_l]$, $\mathbbm{1}_{A} (u)$ is an indicator function with $\mathbbm{1}_{A} (u) = 1$ only if $ u \in A $, and 
$\mathbbm{1}_{ A } (u) = 0$ otherwise. 
Then, the PDF can be expressed as follows: 
\be
f_{h_p} \left( h_p \right) = F_u(a_l) \delta (h_p-1) + (1-F_u(a_l)) \delta(h_p),
\label{eq:f_hp_indic}
\ee
where $\delta(\cdot)$ is the impulse function, and $F_u(\cdot)$ is the cumulative probability distribution~(CDF) of displacement which is a Rayleigh random variable. 
Note that \eqref{eq:f_hp_indic} can be integrated.  

\begin{figure}
\centering
  \includegraphics[width=\userdefinelength]{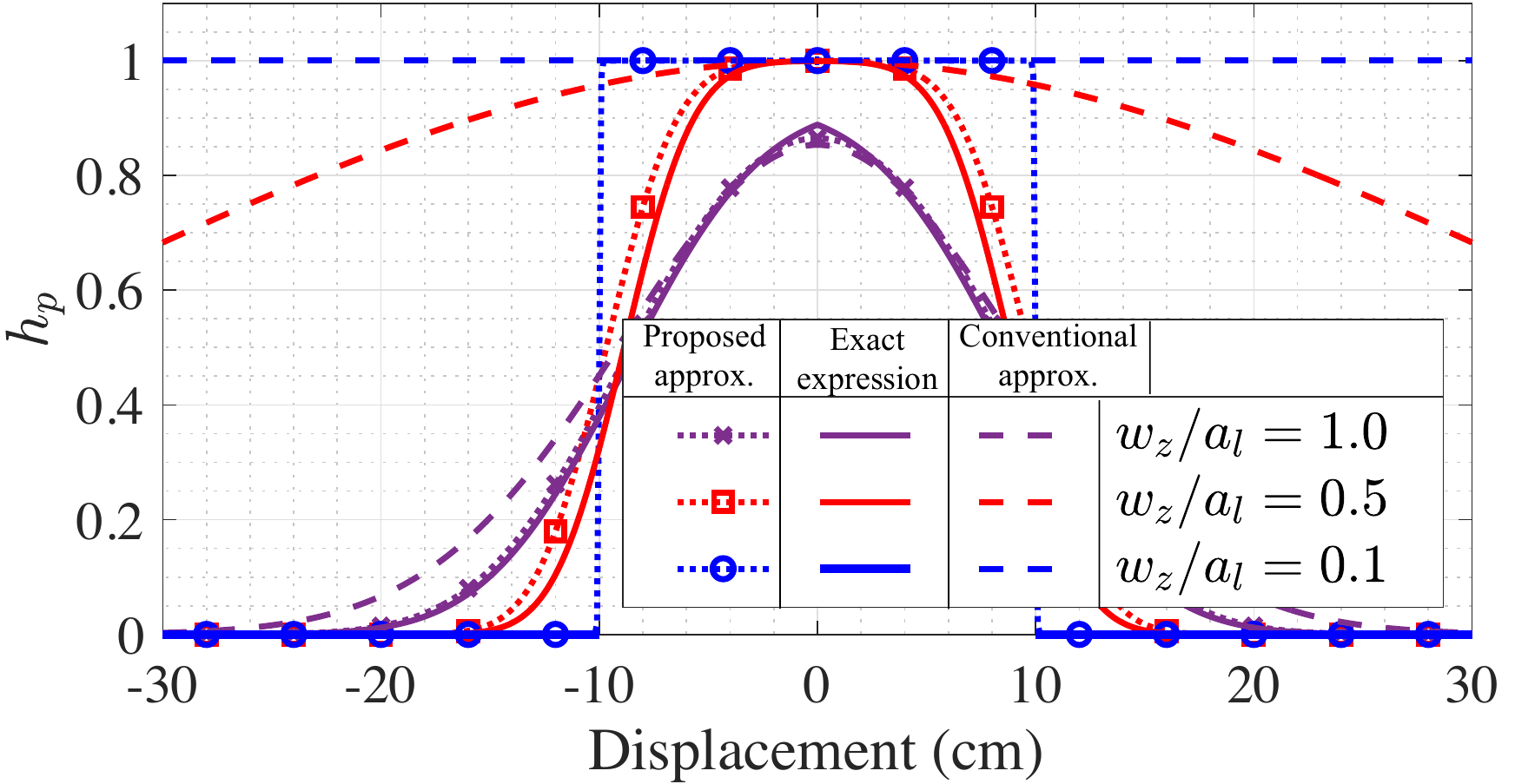}
  \captionof{figure}{Comparison of $h_p(u;w_l,a_l)$ modeling for different values of $w(z)/a_l$.}
  \label{fig:comparison_hp}
  \vspace{-15pt}
\end{figure}

The exact expression, \eqref{eq:exact_h_p}, and proposed approximation model for $h_p(r)$ are plotted in Fig. \ref{fig:comparison_hp} for the three case of $w_l/a_l \in \{0.1, 0.5, 1.0\}$. 
For the approximate model of $h_p$, \eqref{eq:erf_appro_h_p} is used when $w_l/a_l \in \{0.5, 1.0\}$, and \eqref{eq:Indic_hp} is used when $w_l/a_l = 0.1$. 
In addition, we compared the proposed approximation models of $h_p$ with conventional Gaussian form modeling \cite{Farid:2007}. 
Note that the exact expression in \eqref{eq:exact_h_p} and the proposed approximation model correspond for all three cases. 
However, we can observe that the conventional approximation deviates from the exact expression as $w_l/a_l$ decreases.

\subsection{Outage Probability}

In this section, we obtain the outage probability when considering atmospheric and pointing error losses together with the quality of the FSO channel. 
Given $f_{h_p} (h_p)$, the PDF of the composite channel coefficient $h$ is expressed as 
\ben
f_h(h) &=& \int f_{h_p} \left( h_p  \right) f_{h|h_p} \left( h|h_p \right) dh_p \\
&\stackrel{(a)}{=}& \int^{A_0}_{0} f_{h_p} \left( h_p  \right) \frac{1}{h_p h_l} f_{h_a} \left( \frac{h}{h_p h_l} \right) dh_p 
\label{eq:f_h}
\een
where (a) is derived from $h = h_l h_p h_a$. 
We can then compute the outage probability as 
\be
P_{out} = \int^{h_0}_{0} f_h (h) dh.
\label{eq:Pout}
\ee

By substituting $f_h(h)$ from \eqref{eq:f_h}
into \eqref{eq:Pout} and changing the order of integration, the outage probability becomes 
\ben
P_{out} &=& \int^{A_0}_{0} \frac{1}{h_p h_l} f_{h_p}\left( h_p  \right)  \int^{h_0}_0    f_{h_a} \left( \frac{h}{h_p h_l} \right) dh \, dh_p \nonumber \\
&=& \int^{A_0}_{0} f_{h_p}\left( h_p  \right) F_{h_a} \left( \frac{h_0}{h_p h_l}  \right) dh_p
\label{eq:Pout_expand}
\een
where $F_{h_a} (\cdot)$ is the CDF of turbulence-induced channel $h_a$. Note that we consider both log-normal fading for weak turbulence channel and Gamma-Gamma fading for strong turbulence channel. 

For the lognormal fading, we have 
\be
F_{h_a} \left( h_a  \right) = 
\frac{1}{2} + \frac{1}{2} \mbox{erf} 
\left( \frac{ \log \left( h_a \right) }{\sqrt{8 \sigma^2}} \right), 
\label{eq:cdf_lognormal}
\ee
where $\sigma^2\approx \sigma_R^2/4$ is the variance of the log fading amplitude, that is, $0.5 \log(h_a)$. 
$\sigma_R^2$ is the Rytov variance, which is given by 
$\sigma_R^2= 1.23  C_n^2 k^{7/6} d_{e2e}^{11/6}$ where $C_n^2$ is the index of refraction structure parameter. 
For the Gamma-Gamma fading, we have
\begin{multline}
F_{h_a} \left( h_a  \right) = 
\pi \mbox{csc}\left( \pi \gamma  \right) 
\left[ \frac{\left( \alpha \beta h_a  \right)^{\beta} \pFq{1}{2}(\beta;1-\gamma,1+\beta;\alpha\beta h_a)}{\Gamma(\alpha)\Gamma(1-\gamma)\Gamma(\beta+1)} \right. \\
- \left. \frac{\left( \alpha \beta h_a  \right)^{\alpha} \pFq{1}{2}(\alpha;1+\gamma,1+\alpha;\alpha\beta h_a)}{\Gamma(\beta)\Gamma(1+\gamma)\Gamma(\alpha+1)} \right], 
\label{eq:cdf_gg}
\end{multline}
where 
\begin{align} \alpha=&\left [{\exp \left({\tfrac {0.49\sigma _{R}^{2}}{(1+1.11\sigma _{R}^{12/5})^{7/6}}}\right)-1}\right]^{-1}\text {and} \\ \beta=&\left [{\exp \left({\tfrac {0.51\sigma _{R}^{2}}{(1+0.69\sigma _{R}^{12/5})^{5/6}}}\right)-1}\right]^{-1},
\end{align}
$\gamma = \alpha - \beta$, and 
$\pFq{p}{q}$ is the generalized hypergeometric function for integers $p$ and $q$ \cite{Gradshteyn:2007}.

Then, the outage probability can be obtained by substituting $f_{h_p}(h_p)$ from \eqref{eq:hp_pdf_erf} and \eqref{eq:f_hp_indic}, and $F_{h_a}(h_a)$ from \eqref{eq:cdf_lognormal} and \eqref{eq:cdf_gg} into \eqref{eq:Pout_expand}, according to the pointing error model and turbulence-induced fading models. 
Note that for \eqref{eq:hp_pdf_erf}, the outage probability cannot be computed in closed form. However, it involves only one finite integral that can be easily computed numerically. 

Meanwhile, for \eqref{eq:f_hp_indic}, the outage probability can be obtained in closed form, and it is given by
\ben
P_{out} &=& F_u(a_l) \int^{1}_{0} \delta (h_p-1) F_{h_a} \left( \frac{h_0}{h_ph_l}\right) dh_p \nonumber \\
 &&+\left(1-F_u(a_l) \right) \int^{1}_{0} \delta(h_p) F_{h_a} \left( \frac{h_0}{h_p h_l}  \right) dh_p \nonumber \\
 &=& F_u(a_l) \left( F_{h_a} \left( \frac{h_0}{h_l} \right) -1 \right) + 1.
\label{eq:p_out_small_w}
\een
Moreover, from \eqref{eq:p_out_small_w}, 
we can also identify that the outage probability reaches a lower limit when $P_t$ approaches infinity as follows:
\begin{multline}
P_{out} \geq \lim_{P_t \to \infty} P_{out} \\
= \lim_{h_0 \to 0} 
F_r(a) \left( F_{h_a} \left( \frac{h_0}{h_l} \right) -1 \right) + 1 = 1 - F_r(a).   
\label{eq:pout_lbound}
\end{multline}
Note that even if the power increases, the outage probability does not decrease accordingly and gradually converges to the lower limit. In addition, we observe that this lower limit of the outage probability depends only on the probabilistic characteristics of the pointing error, regardless of the atmospheric effect. 

\vspace{-15pt}
\section{Numerical Results}
\label{sec:sim}
\begin{figure}
\centering
  \includegraphics[width=0.9\linewidth]{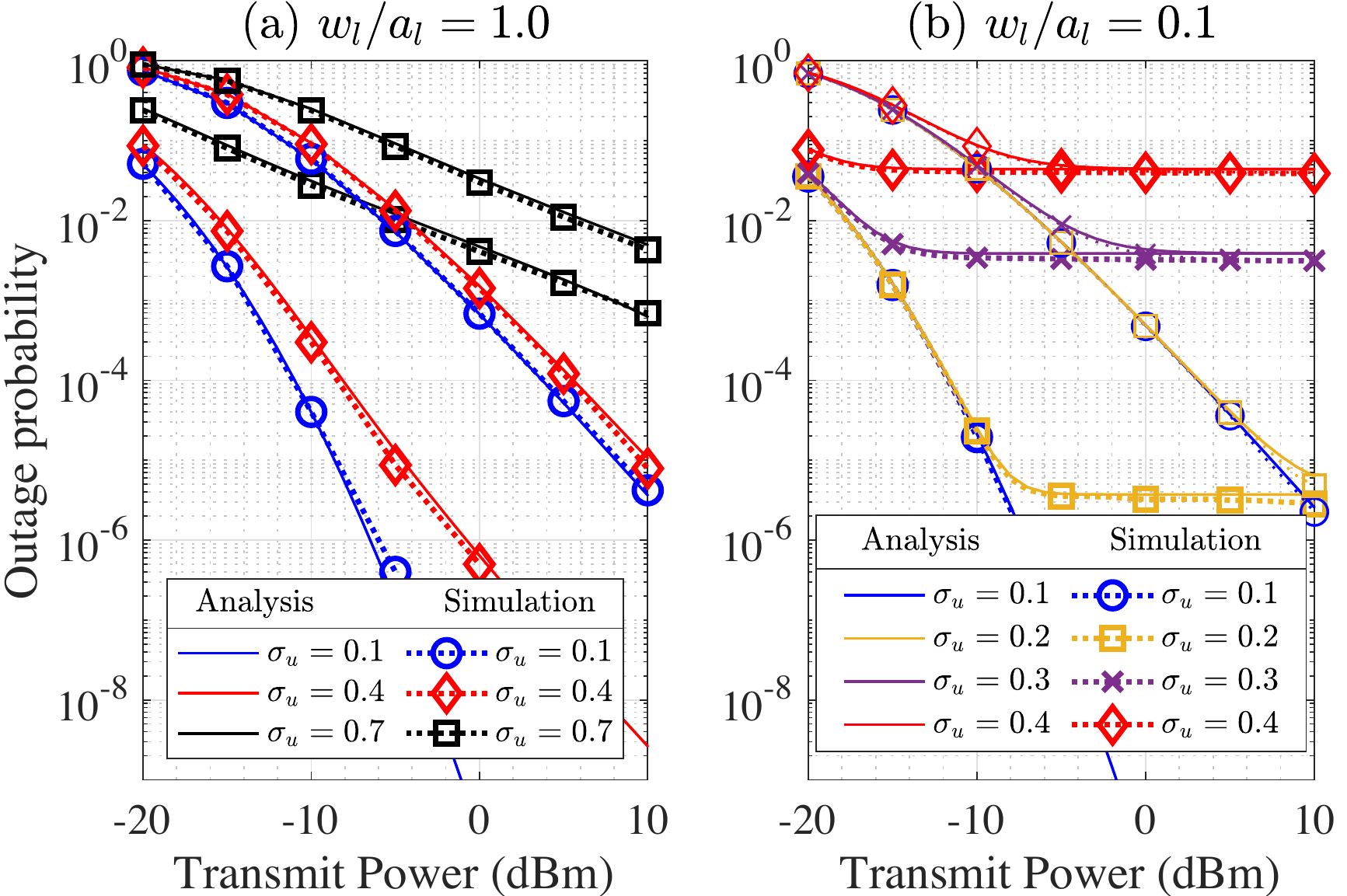}
  \captionof{figure}{Outage probability comparison between analysis and simulation}
  \label{fig:pout_comparison}
  \vspace{-20pt}
\end{figure}
We performed Monte Carlo simulation to verify the proposed pointing error model and the outage probability. 
The simulation parameters are as follows. 
For all simulation cases, we assumed that 
$\lambda = 1550$~nm, and the total length of the propagation path between the Tx to Rx via IRS, $d_{t_2l}$, was 1 km.

The weather conditions on FSO link are characterized by two cases: clear weather with weak turbulence, and light fog with strong turbulence, as in \cite{Farid:2007}. 
For the clear whether, we assumed that the visibility was $10$ km; thus $h_l=0.9$. In the light foggy whether, the visibility was assumed to be $0.5$~km, and $h_l=0.08$. 
The strength of turbulence was characterized by $C_n^2$ and $\sigma_R^2$. 
We assumed that $C_n^2 = \num{5e-14}$ and $\sigma_R^2 = 1$ for the weak turbulence, and  
$C_n^2 = \num{0.5e-14}$ and $\sigma_R^2 = 0.1$ for the strong turbulence.

Fig. \ref{fig:pout_comparison} compares the outage probability of the analysis and simulation according to the standard deviation of the beam's displacement, $\sigma_u$, and transmitted power $P_t$. 
The simulation and analytical results match closely for both the proposed pointing error models, \eqref{eq:erf_appro_h_p} and \eqref{eq:Indic_hp}. 
In addition, we observe that the outage probability reaches the lower limit, which increases with $\sigma_u$.
Moreover, the lower limit of the outage probability was equivalent for both atmospheric models. 
Thus, \eqref{eq:pout_lbound} was verified. 

\begin{figure}
\centering
  \includegraphics[width=0.9\linewidth]{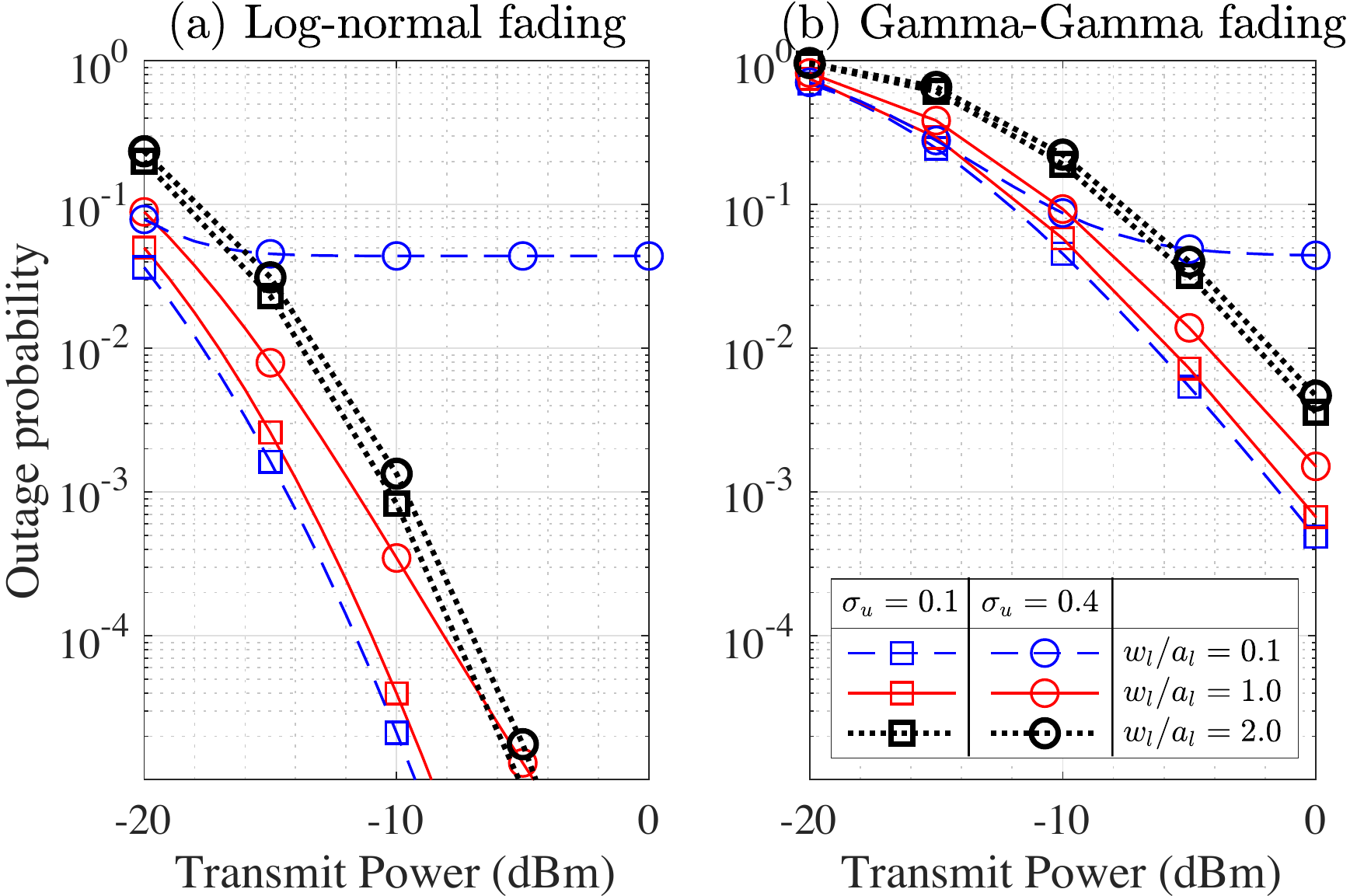}
  \captionof{figure}{Outage probability according to $w_l/a_l$ and $\sigma_u$. Note that (13), (11), and conventional Gaussian form are used for each case of $w_l/a_l \in \{0.1, 1.0, 2.0 \}$, respectively.}
  \label{fig:pout_comparison2}
  \vspace{-15pt}
\end{figure}

Fig. \ref{fig:pout_comparison2} shows that how the outage probability varies according to the beam width and pointing error. 
We can obtain a better outage performance (lower outage probability) with a more focused beam with a small $w_l/a_l$ when the pointing error is marginal, with $\sigma_u=0.1$. 
However, when the pointing error is significant, the more focused beam gets the worse outage performance (high outage probability); 
when $w_l/a_l=0.1$ and $\sigma_u=0.4$, the outage probability reaches the lower bound early about at a power of $-15$ dBm and is larger than other cases of $w_l/a_l \in \{1.0, 2.0\}$ for the most of transmitted power ranges.

Furthermore, we can observe that the performance gain of $w_l/a=0.1$ compared to $w_l/a=1$ is less than $1$ dB when $\sigma_u=0.1$.
Based on this observation, 
it can be deduced that a focused beam smaller than size of the Rx has no substantial benefit.
We can transmit sufficient power with beam whose beam width at the Rx is similar to the radius of the Rx aperture. 
If the beam width is further reduced, it becomes more sensitive to pointing error and only increases the impact of the pointing error. 

\section{Conclusion}
\label{sec:con}

In this paper, we proposed a phase-shift design of an IRS for focusing beams. In addition, we introduced a new pointing error model and an outage probability expression, applicable when the beam width is less than the Rx aperture size, for cases where the conventional model cannot be applied. The simulation validated the proposed phase-shift design, and the outage probability expression was verified. The mathematical modeling presented in this work can facilitate performance analysis according to various factors, such as beam width, power, and pointing error for the focused beam. Thus, this study provides essential preliminary research results for future studies that assume a focused beam in FSO systems.

\bibliographystyle{IEEEtran}

\end{document}